\documentstyle[12pt]{article}
\def\three_j(#1,#2,#3,#4,#5,#6){\pmatrix{#1 & #2 & #3\cr
                                         #4 & #5 & #6\cr}}

\def\qqq{\end{document}}
\def\pmb#1{\setbox0=\hbox{$#1$}%
\kern-.025em\copy0\kern-\wd0
\kern.05em\copy0\kern-\wd0
\kern-.025em\raise.0433em\box0 }

\def\xara(#1,#2,#3,#4){\left(\matrix{#1 & #2\cr #3 & #4\cr}\right)}

\def\six_j(#1,#2,#3,#4,#5,#6){\left\{\matrix{#1 & #2 & #3\cr
                                         #4 & #5 & #6\cr}\right\}}
\def\nine_j(#1,#2,#3,#4,#5,#6,#7,#8,#9){\left\{\matrix{#1 & #2 & #3\cr
                                        #4 & #5 & #6\cr
                                         #7 & #8 & #9\cr}\right\}}

\def\Ener(#1,#2){ \sqrt{{#1}^2+{#2}^2} }

\def\overlay#1#2{\setbox0=\hbox{$#1$}\setbox1=\hbox to \wd0{\hss$#2$\hss}#1%
\hskip -1\wd0\copy1}

\def\bold#1{\setbox0=\hbox{$#1$}%
      \kern-.025em\copy0\kern-\wd0
      \kern.05em\copy0\kern-\wd0
      \kern-.025em\raise.0433em\box0 }

\def\S11{S_{11}(1535)}

\def\footnoterule{\kern-3pt \hrule width \hsize \kern2.6pt}

\newcommand{\be}{\begin{equation}}
\newcommand{\ee}{\end{equation}}
\newcommand{\ba}{\begin{eqnarray}}
\newcommand{\ea}{\end{eqnarray}}

\newcommand{\np}{{\bf p}}

\newcommand{\nne}{{\bf e}}

\newcommand{\nk}{{\bf k}}
\newcommand{\nq}{{\bf q}}
\newcommand{\nkappa}{\mbox{\boldmath $\kappa$}}
\newcommand{\nsigma}{\mbox{\boldmath $\sigma$}}
\newcommand{\neta}{\mbox{\boldmath $\eta$}}

\newcommand{\qbar}{\not{\!Q}}

\newcommand{\md}{m_\Delta}

\newcommand{\nJ}{{\bf      J}}

\newcommand{\nS}{{\bf      S}}

\begin{document}

\begin{titlepage}
\mbox{} 
\vspace*{2.5\fill} 

{\Large\bf 
\begin{center}
%
Relativistic effects in electromagnetic nuclear responses in the
             quasi-elastic delta region 
%
\end{center}
} 

\vspace{1\fill} 

\begin{center}
{\large 
J.E. Amaro$    ^{1}$, 
M.B. Barbaro$  ^{2}$, 
J.A. Caballero$^{3,4}$, 
T.W. Donnelly$ ^{5}$ and 
A. Molinari$   ^{2}$
}
\end{center}

\begin{small}
\begin{center}
$^{1}${\sl 
Departamento de F\'\i sica Moderna,
Universidad de Granada, 
E-18071 Granada, SPAIN 
}\\[2mm]
$^{2}${\sl 
Dipartimento di Fisica Teorica,
Universit\`a di Torino and
INFN, Sezione di Torino \\
Via P. Giuria 1, 10125 Torino, ITALY 
}\\[2mm]
$^{3}${\sl 
Departamento de F\'\i sica At\'omica, Molecular y Nuclear \\ 
Universidad de Sevilla, Apdo. 1065, E-41080 Sevilla, SPAIN 
}\\[2mm]
$^{4}${\sl 
Instituto de Estructura de la Materia, CSIC 
Serrano 123, E-28006 Madrid, SPAIN 
}\\[2mm]
$^{5}${\sl 
Center for Theoretical Physics, Laboratory for Nuclear Science 
and Department of Physics\\
Massachusetts Institute of Technology
Cambridge, MA 02139, USA 
}
\end{center}
\end{small}

\kern 1. cm \hrule \kern 3mm 

\begin{small}\noindent
{\bf Abstract} 
\vspace{3mm} 

A new non-relativistic expansion in terms of the nucleon's
momentum inside nuclear matter of the current for isobar electro-excitation 
from the nucleon is performed. Being 
exact with respect to the transferred energy
and momentum, this yields new current operators which retain
important aspects of relativity not taken into account in the traditional 
non-relativistic reductions. 
The transition current thus obtained differs from
the leading order of the traditional expansion by simple
multiplicative factors. 
These depend on the momentum and energy transfer and can be easily
included together with relativistic kinematics in non-relativistic, 
many-body models of isobar electro-excitation in nuclei.
The merits of the new current are tested by comparing with the
unexpanded electromagnetic nuclear responses in the isobar peak
computed in a relativistic Fermi gas framework. 
The sensitivity of the relativistic responses to the isobar's 
magnetic, electric and Coulomb form factors and the finite width of
the isobar is analyzed.
\kern 2mm 


\end{small}

\kern 2mm \hrule \kern 1cm

\noindent MIT/CTP\#2856 \hfill April 1999

\end{titlepage}


\section{Introduction}


The cross section for inclusive electron scattering $(e,e')$
shows a pronounced peak at an energy transfer 
$\omega \sim \sqrt{q^2+m_N^2}-m_N$, corresponding to the
quasi-free interaction with the individual nucleons in the nucleus 
(here $m_N=$ nucleon mass). 
For high values of the momentum transfer $q=|{\bf q}|$ and higher energy 
loss it is possible to  produce real pions and the cross section
shows another peak dominated by the resonant production
of a $\Delta(1232)$ at
$\omega \sim \sqrt{q^2+m_\Delta^2}-m_N^2$, where $m_\Delta$ is the $\Delta$
mass \cite{Van81}.
The width of these peaks is related to the Fermi momentum of the 
nucleons inside the nucleus and, in the case of the $\Delta$-peak,
also to the decay width of the $\Delta$ in nuclear matter.
Hence for a high enough value of $q$, these two peaks actually 
overlap and cannot be separated in inclusive experiments
\cite{Ang96,Day93}. Thus the response in the region above the
quasi-elastic peak contains information about the nucleon's excited
states and their change due to the nuclear medium. 

Since the electro-excitation of the $\Delta$ requires high energy
and momentum transfers, a relativistic treatment of the reaction is
needed. Recently, several  many-body  calculations, both in nuclear
matter and finite nuclei,  
have been performed in this region~\cite{Gil97,Gad98,Bau98}; all of
these calculations 
are non-relativistic in nature, although some relativistic
corrections enter in two of them, including an expansion of
the current to order $(p/m_N)^2$ in \cite{Gil97} and 
relativistic kinematics in \cite{Gad98}. 
Clearly some of these corrections are inadequate when one wishes 
to go to high momentum transfers, $q\simeq 1$ GeV/c, and there
relativistic models such as the ones developed  in 
\cite{Mon69,Weh89,Cha93,Dek91} are
more appropriate.
However, although these last calculations are fully relativistic, they
do not include the full $N$-$\Delta$
vertex. For instance, in the pioneering calculation by Moniz \cite{Mon69}
the Peccei Lagrangian was used, which is only appropriate 
for computing the transverse response for low momentum transfer
\cite{Duf68,Jon73,Pet98}. In other work \cite{Weh89,Cha93} a more
appropriate M1 magnetic transition current was used, although
the electric E2 and Coulomb C2 excitation amplitudes of the $\Delta$
were not included.
For years an important program has been pursued to determine 
more accurately 
the quadrupole $C2$ and $E2$ amplitudes in the $\Delta$ region 
\cite{Dav86},
these being small compared with the dominant dipole $M1$
amplitude.
Using polarized photons, the E2/M1-ratio has been measured 
to be around  -3\% at resonance \cite{Bec97}. 
The C2 amplitude however only appears in 
electro-production reactions $N(e,e')\Delta$. 
Values of  the ratio $C2/M1$ around $-13\%$ have been reported 
in $H(e,e'\pi^0)p$ experiments at $Q^2=0.13$ (GeV/c)$^2$ \cite{Kal97}.  
However, this value differs with the findings of
recent measurements of the transverse-longitudinal 
asymmetry and proton polarization 
in $H(e,e'p)\pi^0$ reactions \cite{Mer99}.
Hence, for the $N\to\Delta$ transition our knowledge is still
incomplete and has not been possible to undertake a full analysis
in the  sense of the work by Nozawa and Lee \cite{Noz90}
of the effect of the C2 and E2 form factors in the nuclear
$\Delta$-peak.

In this paper we perform a new non-relativistic expansion of the 
electro-excitation current of the nucleon, providing an extension of
our previous expansion of the electromagnetic nucleon
current in powers of $\eta=p/m_N$, in which we retained the full
dependence on 
$q$ and $\omega$ \cite{Ama96}. Recently the same procedure has been also applied
to meson-exchange currents (MEC) \cite{Ama98}.
These currents can be implemented together with relativistic
kinematics in standard non-relativistic models
of one-particle emission near the quasi-elastic peak. 
In this paper we apply the same procedure to the $\Delta$
electro-excitation current, which we develop to leading order in 
$\eta$, again retaining the full dependence on the energy
and momentum transfers. 
We perform this expansion for the magnetic transition current M1,
which is the dominant one both in the longitudinal and transverse
nuclear responses \cite{Cha93}. The resulting current is designed
in such a way that it differs from the traditional
non-relativistic limit simply by having $(q,\omega)$-dependent factors
which multiply the traditional operators. These corrections, being  
of leading order in $\eta$, are seen to arise mainly from the normalization 
factor in the Rarita-Schwinger spinor and from the lower-component
spinology now included in the effective current operator.

The organization of the work is as follows.
In sect.~2 we first develop the analytical expressions for the 
longitudinal and transverse responses in the relativistic Fermi gas
(RFG) model by using a $\Delta$-hole approach. We consider the 
full vertex of Jones and Scadron \cite{Jon73} that includes
M1, E2 and C2 $\Delta$-amplitudes. 
In deriving expressions for these responses we assume a stable 
$\Delta$ particle and
later include its finite width by performing a convolution 
of these responses with a Lorentz distribution. 
In sect.~3 we perform the expansion of the magnetic current 
to leading order in the momentum of the bound nucleon. 
In sect.~4 we test the validity of the expansion by comparing
the exact RFG result with a non-relativistic Fermi gas model,
using the new current and relativistic kinematics. 
Furthermore, we compare several models of the reaction by using the
RFG;
in particular, we study the differences that arise upon using
the Peccei and magnetic Lagrangians,
and explore the effects of the E2, C2 multipoles and the finite width 
of the $\Delta$ on
the longitudinal and transverse response functions. 
Finally in sect.~5 we draw our main conclusions.


\section{General formalism}


We start our discussion by introducing the general formalism, which is
based on a 
relativistic treatment of the nuclear currents entering into the calculation 
of the response functions. 
It is well-known \cite{Alb88} that the longitudinal and transverse 
response functions can be evaluated as components of the nuclear tensor 
$W_{\mu\nu}$, namely

\ba
R_L(q,\omega) &=& \left(\frac{q^2}{Q^2}\right)^2 
\left[ W_{00} - \frac{\omega}{q} \left( W_{03}+W_{30} \right) + 
\frac{\omega^2}{q^2} W_{33} \right]
= W_{00} 
\label{RL}
\\
R_T(q,\omega) &=& W_{11}+W_{22}
\ ,
\label{RT}
\ea
where $Q^\mu = (\omega,\nq)$ is the space-like four-momentum carried by the 
virtual photon and the gauge invariance has been exploited in 
obtaining Eq.~(\ref{RL}).
We shall compute this nuclear tensor in the RFG 
framework, where nucleons are assumed to move freely inside the system
with relativistic kinematics, hence being on their mass-shell.

\subsection{Nuclear tensor and response functions for inelastic processes}

We first consider the electro-production of a stable
resonance (namely, the $\Delta$ is viewed as a particle on its mass-shell)
in the RFG --- 
in this case analytical expressions for the response functions are
obtained ---
and later on we shall include corrections due to inclusion of the decay
width of the $\Delta$.
The RFG nuclear tensor reads (see, for example, \cite{Alb88})

\ba
W_{\mu\nu} 
= 
\frac{3\pi^2{\cal N}m_N^2}{k_F^3}
\int \frac{d\np}{(2\pi)^3} 
\frac{\theta(k_F-p)}{E(\np) E_\Delta(\np+\nq)}
f_{\mu\nu}(\np,\np+\nq)
\delta[E_\Delta(\np+\nq)-E(\np)-\omega]\ ,
\label{Wmunu}
\ea
where ${\cal N}$ is the number of protons or neutrons in the nucleus. Here
$m_N$ and $\md$ are the masses of the struck nucleon and $\Delta$,
respectively, and
$E(\np)=\sqrt{m_N^2+ p^2}$ and 
$E_\Delta(\np+\nq) = \sqrt{\md^2+(\np+\nq)^2}$
are their corresponding energies;
$k_F$ is the Fermi momentum and
$f_{\mu\nu}$ the inelastic single-nucleon tensor of the 
$N\to\Delta$ transition.

Introducing the standard dimensionless variables

\begin{eqnarray}
& &\neta = \frac{\np}{m_N}\ , \ \ 
\neta_F=\frac{\nk_F}{m_N} \ , \ \ 
\nkappa=\frac{\nq}{2m_N} \ , \ \ 
\\
& & \lambda=\frac{\omega}{2m_N}\ , \ \
\tau=\kappa^2-\lambda^2\ , \ \  
\varepsilon=\sqrt{1+\eta^2}\ , \ \ 
\mu_\Delta=\frac{\md}{m_N}
\label{VAR}
\end{eqnarray}
and performing the angular integration in Eq.~(\ref{Wmunu}) we obtain

\ba
W_{\mu\nu} (\kappa,\lambda)
= 
\frac{3{\cal N}}{8 m \eta_F^3\kappa} 
\int_{\varepsilon_0}^{\varepsilon_F} f_{\mu\nu}(\varepsilon,\theta_0) 
d\varepsilon \ ,
\label{Wint}
\ea
where $\varepsilon_F=\sqrt{1+\eta_F^2}$ is the Fermi energy and

\ba
\varepsilon_0 = \kappa \sqrt{\frac{1}{\tau}+\rho^2}
-\lambda\rho
\label{eps0}
\ea
the minimum energy of the struck nucleon for fixed $\kappa$
and $\lambda$, having defined the factor

\ba
\rho = 1 + \frac{1}{4\tau} \left(\mu_\Delta^2-1\right)
\ ,
\ea
which measures the inelasticity of the elementary process.

In Eq.~(\ref{Wint}) the single-nucleon tensor $f_{\mu\nu}(\varepsilon,\theta_0)$
contains the angle $\theta$ between $\neta$ and $\nkappa$ given via

\ba
\cos\theta_0 = 
\frac{\lambda\varepsilon-\tau\rho}{\kappa\eta} \ ,
\label{cost0}
\ea
as required by energy conservation.
The condition $|\cos\theta_0|\leq 1$ then permits the response 
of the $\Delta$ to occur only in the range

\ba
\frac{1}{2} \left[ \sqrt{(2\kappa-\eta_F)^2+\mu_\Delta^2}-\varepsilon_F \right]
\leq \lambda \leq
\frac{1}{2} \left[ \sqrt{(2\kappa+\eta_F)^2+\mu_\Delta^2}-\varepsilon_F \right]
\ .
\label{region}
\ea

In analogy with the physics of the quasi-elastic
peak \cite{Alb88}, it is convenient to introduce a scaling variable 
$\psi_\Delta$ defined as follows

\ba
\psi_\Delta^2(\kappa,\lambda) &=& \frac{\varepsilon_0-1}{\xi_F} 
= \frac{1}{\xi_F}
\left( \kappa \sqrt{\frac{1}{\tau}+\rho^2}
-\lambda\rho - 1\right) \ , 
\label{psi2D}
\ea
with $\xi_F=\varepsilon_F-1$.
The physical meaning of the scaling variable may be deduced from the
above equation: in terms of dimensionless variables, 
$\xi_F \psi_\Delta^2$ is the minimum kinetic energy 
required to transform a nucleon inside the nucleus into a $\Delta$ when hit 
by a photon of energy $\lambda$ and momentum $\kappa$.
It is straightforward to check that, when $m_N=\md$ (hence $\rho=1$), the 
ordinary quasi-elastic scaling variable $\psi$ \cite{Alb88}
is recovered. 

In terms of the scaling variable in Eq.~(\ref{psi2D}) the response region 
given by Eq.~(\ref{region}) simply 
reduces to $-1\leq\psi_\Delta\leq 1$. The energy position $\lambda_{\Delta P}$
of the peak of the $\Delta$ response occurs when  $\psi_\Delta$ vanishes, 
namely for

\ba
\psi_\Delta=0 \ \longrightarrow \ 
\lambda_{\Delta P} = \tau\rho \ \ ,
\kappa^2_{\Delta P} = \tau(\tau\rho^2+1) \ .
\label{peak}
\ea

We now turn to a consideration of the nucleonic tensor $f_{\mu\nu}$,
which we will obtain in the next section for specific 
$N\to\Delta$ currents. For any physical process this tensor must 
comply with Lorentz covariance and current conservation 
$(Q^\mu f_{\mu\nu} = f_{\mu\nu}Q^\nu =0)$. The most general
unpolarized second-rank 
tensor consistent with these requirements is \cite{Don92}

\ba
f_{\mu\nu}(\np,\np+\nq) = 
- w_1(\tau) \left( g_{\mu\nu} - \frac{Q_\mu Q_\nu}{Q^2} \right)
+ w_2(\tau) V_\mu V_\nu - \frac{i}{m_N} w_3(\tau) 
\varepsilon_{\mu\nu\rho\sigma}
Q^\rho V^\sigma \ ,
\label{fmunu}
\ea
where  $w_1(\tau)$, $w_2(\tau)$ and
$w_3(\tau)$ are {\em scalar} functions containing the specific
dynamics
of the process and

\ba
V_\mu = \frac{1}{m_N} \left( P_\mu - \frac{P\cdot Q}{Q^2} Q_\mu \right) \ ,
\label{Vmu}
\ea
is a four-vector orthogonal to $Q_\mu$, $P_\mu$ being the struck nucleon's 
four-momentum. Only the terms involving $w_{1,2}(\tau)$ occur in
EM interactions, whereas $w_3(\tau)$ also enters for the full
electroweak interaction. Energy conservation 
via Eq.~(\ref{cost0}) implies that $P\cdot Q/Q^2=-\rho/2$: hence
the longitudinal and transverse components of the single-nucleon tensor read

\ba
f_L =
f_{00} = 
-\frac{\kappa^2}{\tau} w_1(\tau) + 
(\lambda\rho+\varepsilon)^2 w_2(\tau) 
\label{fL}
\ea
and 

\ba
f_T =
f_{11} + f_{22} =
2 w_1(\tau) 
+ \left[\varepsilon^2-1 - \left(\frac{\lambda\varepsilon-
\tau\rho}{\kappa}\right)^2 \right] w_2(\tau) \ .
\label{fT}
\ea

Finally, by performing the energy integral in Eq.~(\ref{Wint}) one gets for the 
response functions in Eqs.~({\ref{RL},\ref{RT}) the following expressions

\ba
R_L(\kappa,\lambda) &=&
\frac{3{\cal N} \xi_F}{8 m_N\eta_F^3\kappa} 
\frac{\kappa^2}{\tau} \left[(1+\tau\rho^2) w_2(\tau) 
- w_1(\tau) + w_2(\tau) {\cal D}(\kappa,\lambda) \right]
(1-\psi_\Delta^2) \theta(1-\psi_\Delta^2)
\nonumber\\
&&
\label{RLRFG}
\\
R_T(\kappa,\lambda) &=&
\frac{3{\cal N} \xi_F}{8 m_N\eta_F^3\kappa} 
\left[2 w_1(\tau) + w_2(\tau) {\cal D}(\kappa,\lambda) \right] 
(1-\psi_\Delta^2) \theta(1-\psi_\Delta^2) \ ,
\label{RTRFG}
\ea
where 

\ba
{\cal D}(\kappa,\lambda) &\equiv& 
\frac{\tau}{\kappa^2} \left[
(\lambda\rho+1)^2 + (\lambda\rho+1) (1+\psi_\Delta^2) \xi_F 
+ \frac{1}{3} (1+\psi_\Delta^2+\psi_\Delta^4) \xi_F^2 \right] 
-(1 +\tau \rho^2)
\nonumber\\
&&
\label{Delta}
\ea
reflects the (modest) Fermi motion of the nucleons.
In fact, at the resonance peak $(\psi_\Delta=0)$ Eq.~(\ref{Delta}) reduces to

\ba
{\cal D}(\kappa,\lambda)_{\Delta P} = \xi_F + \frac{\tau}{3\kappa^2} \xi_F^2 \ ,
\ea
and, since $\xi_F\simeq 0.03$, yields a small correction.

\subsection{Density dependence of the response functions}

In this subsection we briefly explore the density 
dependence of the previously deduced responses.
This is conveniently achieved by performing an expansion in the parameter 
$\xi_F$. The leading terms of the expansion of

\be
R_{L,T}(\kappa,\lambda) =
\frac{3{\cal N} \xi_F}{8 m\eta_F^3\kappa} (1-\psi_\Delta^2)
\frac{\kappa^2}{\tau} 
\left( R_{L,T}^{(0)} + R_{L,T}^{(1)} \xi_F + R_{L,T}^{(2)} \xi_F^2 \right)
\ee
are given by
\ba
R_L^{(0)} &=& 
- w_1(\tau) + \frac{\tau}{\kappa^2} 
\left( \lambda \rho + 1 \right)^2 w_2(\tau) 
\\
R_T^{(0)} &=& 
2 w_1(\tau) + \left\{
-\left(1 +\tau \rho^2 \right)
+ \frac{\tau}{\kappa^2} 
\left( \lambda \rho +1 \right)^2 \right\} w_2(\tau) 
\ ,
\ea
and the next terms are found to be 
\ba
R_L^{(1)} &=& R_T^{(1)} =
\frac{\tau}{\kappa^2} 
\left( \lambda \rho + 1 \right) 
(1+\psi_\Delta^2) w_2(\tau) 
\\
R_L^{(2)} &=& R_T^{(2)} =
\frac{1}{3} \frac{\tau}{\kappa^2} (1+\psi_\Delta^2+\psi_\Delta^4) w_2(\tau)
\ .
\ea

In performing the $\xi_F\to 0$ limit it is of importance to realize that both
responses shrink to the peak where $\kappa^2=\tau(\tau\rho^2+1)$ and 
$\lambda=\tau\rho$. This 
constraint requires that ${\cal D}\to 0$ when $\xi_F\to 0$ (see
Eq.~(\ref{Delta})) and hence

\ba
R_L^{(0)}(\kappa,\lambda;\xi_F=0) &=& -w_1(\tau)+(1+\tau\rho^2) w_2(\tau) 
\label{RLxif0}
\\
R_T^{(0)}(\kappa,\lambda;\xi_F=0) &=& 2 w_1(\tau) \ .
\label{RTxif0}
\ea
Expressions for
the $\Delta$ responses in this limit will be given later. In the nucleonic
sector one immediately obtains $R_L^{(0)}=G_E^2$ and $R_T^{(0)}=2\tau G_M^2$
 (cf. \cite{Alb88}).

\smallskip

The expressions in Eqs.~(\ref{RLRFG}), (\ref{RTRFG}) and  (\ref{Delta})
are valid for any process involving an initial nucleon which is converted to an
on-shell resonance of mass $\md$. In particular, the limit $\rho=1$ yields the 
response functions in the quasi-elastic peak region \cite{Alb88}.
The specific physical process gives rise to different $w_1$ and $w_2$ 
functions, which are evaluated in the next section for the
$N\to\Delta$ transition.

\subsection{Nucleonic tensor}

We now evaluate the invariant functions $w_1$ and $w_2$ relative to the
$\gamma N\to\Delta$ process. These functions, being scalars, can be computed
in any reference system and the most convenient one
is found to be the rest system of the $\Delta$, where the Rarita-Schwinger
spinors take their simplest form.  
Let us denote with $Q^*_\mu=(\omega^*,\nq^*)$ the 
four-momentum transfer in this system
(the corresponding four-vector in the nucleus laboratory frame is $Q_\mu$) 
and let $P^*_\mu=(E^*,\np^*)$ be the four-momentum of the struck nucleon.
The $\Delta$ system is then defined by
\begin{equation}
\np^*_\Delta=0 \Rightarrow \np^* =
-\nq^* =(0,0,- q^*)\ ,
\end{equation}
having chosen the $z$-axis in the direction of $\nq^*$.
The energy conservation condition accordingly reads
\begin{equation}
E^*= \md -\omega^* \ .
\end{equation}
Using the energy-momentum relation for the initial nucleon, namely
\begin{equation}
m_N^2 = E^{*2}- p^{*2} 
= \md^2-2\md \omega^*+ Q^2 \ ,
\label{Mass}
\end{equation}
we find the value of $\omega^*$ in terms of $Q^2$

\begin{equation}\label{omega}
\omega^* = \frac{\md^2-m_N^2+Q^2}{2\md} \ .
\end{equation}
In this system $V_\mu$ has its space components parallel to $\nq^*$, 
while its time component reads

\begin{equation}
V^*_0  = \frac{1}{m_N}\left( E^*-
\frac{P\cdot Q}{Q^2}\omega^*\right)
     =  -\frac{\md q^{*2}}{m_N Q^2} \ .
\end{equation}

Then we can easily compute the 00 and 11 components of the nucleonic tensor
in Eq.~(\ref{fmunu}), obtaining:

\begin{eqnarray}
 f^*_{11} & = & -w_1 g_{11} = w_1 \\
 f^*_{00} & = & -w_1\left(1-\frac{\omega^{*2}}{Q^2}\right)
             +w_2\frac{\md^2 q^{*4}}{m_N^2 Q^4}
             \nonumber\\
       & = & w_1\frac{q^{*2}}{Q^2}
            +w_2\frac{\md^2 q^{*4}}{m_N^2  Q^4} \ .
\end{eqnarray}
By inverting the above equations the structure functions in the 
$\Delta$-system are found to be

\begin{eqnarray}
w_1 &=& f^*_{11} 
\label{w1f}\\
w_2 &=& \frac{m_N^2 Q^4}{\md^2 q^{*4}}
        \left( f^*_{00}-\frac{q^{*2}}{Q^2} f^*_{11}\right) \ .
\label{w2f}
\end{eqnarray}
Hence the problem is reduced to computing just two components, namely
$f^*_{00}$ and $f^*_{11}$, of the nucleon tensor.

With Jones \& Scadron \cite{Jon73} we write
the transition matrix element for the $\Delta$ excitation as follows 

\begin{equation}
\langle \Delta | j_\mu | N\rangle
= \overline{u}^\beta\Gamma_{\mu\beta}u \ ,
\label{matrix}
\end{equation}
the most general form of the vertex being
\begin{equation}\label{Gamma}
\Gamma_{\mu\beta} = 
C_1\Gamma^1_{\mu\beta}
+C_2\Gamma^2_{\mu\beta}
+C_3\Gamma^3_{\mu\beta} \ .
\label{JS}
\end{equation}
In the above the $C_a$ are (invariant) form factors and the 
three couplings
$\Gamma^a_{\mu\beta}$ are given by%
\footnote{Note that the expressions for the transition matrix element
of Dufner and Tsai \cite{Duf68} 
and of Devenish et al. \cite{Dev76} differ from the ones of Jones and Scadron
because the latter employ the set of basis vectors $(K,Q)$, whereas Dufner
and Devenish use $(P_\Delta,P)$ and $(P_\Delta,Q)$ respectively.
Hence both $\Gamma_2$ and $\Gamma_3$ and the form factors $C_2$ and $C_3$ 
of these three sets of authors will be different.}
\begin{eqnarray}
\Gamma^1_{\mu\beta}
&=& (Q_\beta\gamma_\mu-\qbar g_{\beta\mu})\gamma_5 T_3^{+}
\label{Gamma-1}\\
\Gamma^2_{\mu\beta}
&=& (Q_\beta K_\mu-Q\cdot K g_{\beta\mu})\gamma_5 T_3^{+}\\
\Gamma^3_{\mu\beta}
&=& (Q_\beta Q_\mu-Q^2 g_{\beta\mu})\gamma_5 T_3^{+} \ ,
\end{eqnarray}
where $K_\mu=(P_\mu+P_{\Delta\mu})/2$ and ${\bf T^{+}}$ is the 
$N\to\Delta$ isospin transition operator \cite{Eri88}.

The $N\to\Delta$ tensor $f^*_{\mu\nu}$ then reads

\begin{equation}
f^*_{\mu\nu} = \frac{4}{3} \frac{\md}{m_N} \sum_{ss_\Delta}
          (\overline{u}_\Delta^\lambda\Gamma_{\mu\lambda}u)^*
          (\overline{u}_\Delta^\beta\Gamma_{\nu\beta}u) \ ,
\end{equation}
the factor $4/3$ arising from the isospin trace 

\begin{equation}
\sum_{tt_\Delta} <t|T_3|t_\Delta><t_\Delta|T_3^{+}|t> = \frac{4}{3}
\end{equation}
and $u_\Delta^\beta$
being the Rarita-Schwinger spinor describing a spin 3/2 particle. 
In the $\Delta$ rest frame the latter has the simple form

\begin{eqnarray}
u^0_\Delta({\bf 0},s_\Delta) &=& 0 \\
u^i_\Delta({\bf 0},s_\Delta) &=& \sum_{\lambda s'}
                  {\textstyle 
                       \langle \frac12 s' 1 \lambda
                       |\frac32s_\Delta\rangle}
                  e_\lambda^i u_\Delta({\bf 0},s') \ ,
\end{eqnarray}
where the $\nne_\lambda$ ($\lambda=-1,0,+1$) are spherical vectors
and
\begin{equation}
u_\Delta({\bf 0},s') = \left(\begin{array}{c} 
                        \chi'_{s'} \\ 
                         0 
                        \end{array}
                 \right)
\end{equation}
is a zero momentum $1/2$-spinor. 

Since the initial nucleon spinor is
\begin{equation}
u (\np^*,s) = \sqrt{\frac{1+\varepsilon^*}{2}}
            \left(\begin{array}{c} 
                   \chi_{s} \\ 
                   \frac{\nsigma\cdot\neta^*}{1+\varepsilon^*}\chi_s
                  \end{array}
            \right) \ ,
\end{equation}
the matrix element of the four-dimensional gamma-matrix 
\begin{equation}
\Gamma = \left( \begin{array}{cc}
         \Gamma_{11} & \Gamma_{12} \\
         \Gamma_{21} & \Gamma_{22}
       \end{array}
\right)
\end{equation}
in the $\Delta$-system will be related to the corresponding bispinor
matrix element according to
\begin{equation}
\overline{u}_\Delta\Gamma u 
 = 
\sqrt{\frac{1+\varepsilon^*}{2}}
\chi'{}^{\dagger}
\left[ \Gamma_{11}
      +\Gamma_{12}\frac{\nsigma\cdot\neta^*}{1+\varepsilon^*}
\right] 
\chi
\equiv \chi'{}^{\dagger}\overline\Gamma \chi \ ,
\end{equation}
which defines $\overline\Gamma$.
The general matrix element between $1/2$ and $3/2$ spinors 
can be written as
\begin{equation}
\overline{u}^i_\Delta\Gamma u 
=  {\textstyle 
        \langle \frac32 s_\Delta|
        S^{i\dagger}
        \overline\Gamma
        |\frac12 s\rangle
     } \ ,
\end{equation}
where the spin transition operator for the $\Delta$
\begin{equation}
     {\textstyle 
        \langle \frac32 s_\Delta|
        (S^{\dagger})_\lambda
        |\frac12s'\rangle
=            \langle \frac12 s' 1 \lambda
           |\frac32s_\Delta\rangle}
\label{SPIN}
\end{equation}
has been introduced.

The tensor $f^*_{\mu\nu}$ evaluated in the $\Delta$ rest frame then turns
out to be

\begin{eqnarray}
f^*_{\mu\nu}
&=& \frac{4}{3} \frac{m_N}{\md}
    \sum_{ss_\Delta} 
        {\textstyle 
        \langle \frac32 s_\Delta|
        S^{i\dagger}
        \overline\Gamma_{\mu i}
        |\frac12 s\rangle^*
        \langle \frac32 s_\Delta|
        S^{j\dagger}
        \overline\Gamma_{\nu j}
        |\frac12 s\rangle
     }
\nonumber\\
&=& \frac{4}{3} \frac{m_N}{\md}
    {\rm Tr}
    \left[ \overline\Gamma_{\mu i}^{\dagger}
        \left(\frac{2}{3}\delta_{ij}-\frac{i}{3}\epsilon_{ijk}\sigma_k\right) 
        \overline\Gamma_{\nu j}
    \right] 
\nonumber\\
&=& \frac{4}{9} \frac{m_N}{\md}
    {\rm Tr}
    \left[ 2\overline\Gamma_{\mu i}^{\dagger}
           \overline\Gamma_{\nu i}
          -i\epsilon_{ijk}
            \overline\Gamma_{\mu i}^{\dagger}
            \sigma_k
            \overline\Gamma_{\nu j}
    \right] \ ,
\end{eqnarray}
where the trace is meant to be performed in a two-dimensional non-relativistic
spin space.

By inserting Eq.~(\ref{JS}) into the above expression and by
 performing the traces we get 

\begin{eqnarray}
f^*_{00} &=& 
\frac{8\md}{9m_N^2}q^{*2}(E^*-m_N) 
\left[ C_1+\frac{E^*+\md}{2}C_2+\omega^* C_3 \right]^2
\label{fst00}
\\
f^*_{11} &=&
\frac{8\md}{9m_N^2}
(E^*-m_N) 
\left[C_1^2 (E^*+m_N)^2+C^2(m_N+\md)^2 -C_1C(m_N+\md)(E^*+m_N)\right]
\ ,
\nonumber\\
\label{fst11}
\end{eqnarray}
where a form factor

\begin{equation}
C =  C_1+\frac{1}{2}C_2(\md-m_N)+C_3 \frac{Q^2}{\md+m_N}
\end{equation}
has been defined.

It thus appears from Eq.~(\ref{fst00}) that $f^*_{00}$ only depends upon a
particular, quadratic 
 combination of the three form factors $C_i$, namely 

\begin{equation}
G_C = \frac{4 m_\Delta m_N}{3(\md+m_N)}
      \left[C_1
            +\frac12(E^*+\md)C_2
            +\omega^* C_3
      \right] \ ,
\label{GC}
\end{equation}
commonly referred to as the Coulomb form factor. 
On the other hand, the expression $f^*_{11}$, in addition to the terms
$C_1^2$ and $C^2$, also contains their cross product
$CC_1$. In the spirit of the familiar Sachs form factors, new form
factors can however be defined in such a way
that only squares will appear: indeed by diagonalizing the quadratic form 
in Eq.~(\ref{fst11}) one obtains

\begin{equation}
f^*_{11} = \frac{8\md}{9m_N^2}(E^*-\md)
\left[\frac{3(m_N+\md)}{4m_N}\right]^2 [G_M^2+3G_E^2] \ ,
\label{f*11}
\end{equation}
where 
\begin{eqnarray}
G_E &=&
     \frac{2}{3} m_N
     \left( C - C_1 \frac{E^*+m_N}{\md+m_N} \right)
\label{GE}
\\
G_M &=& 
     \frac{2}{3} m_N
     \left( C + C_1 \frac{E^*+m_N}{\md+m_N} \right)
\label{GM}
\end{eqnarray}
are usually referred to as electric and magnetic form factors.
It is easily verified that the expressions in Eqs.~(\ref{GC}),
(\ref{GE}) and (\ref{GM}) coincide with the definitions of the form factors 
given in \cite{Jon73}.
Furthermore the combination $G_M^2+3G_E^2$ entering in Eq.~(\ref{f*11}) is the 
one which usually appears in the generalized Rosenbluth formula for the 
transverse cross section in terms of magnetic and electric form 
factors \cite{Jon73}.  

Finally, by means of Eqs.~(\ref{w1f},\ref{w2f}) we obtain the two structure 
functions $w_1$ and $w_2$:

\begin{eqnarray}
w_1 &=& \frac{8\md}{9m_N^2}(E^*-m_N)
\left[\frac{3(m_N+\md)}{4m_N}\right]^2 \left(G_M^2+3G_E^2\right)
\label{w1d}
\\
w_2 &=& \frac{8Q^2}{9\md q^{*2}}(E^*-m_N)
 \left[\frac{3(m_N+\md)}{4m_N}\right]^2
\left(\frac{Q^2}{\md^2}G_C^2 -G_M^2-3G_E^2\right) \ .
\label{w2d}
\end{eqnarray}

To finish this section it is useful to write these structure functions in an
arbitrary system, a result which will be also used
in the next section to compare with the 
non-relativistic result. 
Here we only consider the structure functions coming from the magnetic
$M1$ nucleon multipole (proportional to $G_M^2$),
which is the leading contribution. Later in the 
calculations we will study the effect of including the electric and
Coulomb form factors.
In order to relate these expressions to the ones
obtained using the so-called magnetic form of the operator, 
we introduce a new dimensionless form factor $G$ defined by
\begin{equation}\label{G-form-factor}
G = -\frac{m_N+m_\Delta}{2m_N}\frac{3m_N^2G_M}{(m_N+m_\Delta)^2-Q^2} \ .
\end{equation}
Using this form factor, the magnetic invariant functions can be written
\begin{eqnarray}
w_1 &=& \frac{8m_\Delta^3}{9m_N^6}q^{*2}(E^*+m_N)G^2
\\
w_2 &=& -\frac{m_N^2Q^2}{m_\Delta^2q^{*2}}w_1 \ ,
\end{eqnarray}
where we have used the fact that in the $\Delta$-system
\begin{equation}
(m_N+m_\Delta)^2-Q^2 = 2m_\Delta(E^*+m_N) \ .
\end{equation}
In the $\Delta$ system we have the following expressions for the 
quantities involved, written in terms of the invariant $\tau$:
\begin{eqnarray}
q^{*2}  &=& 4m_N^2 \tau\frac{1+\tau\rho^2}{\mu_\Delta^2}\\
m_N+E^* &=& \frac{m_N}{\mu_\Delta}(1+\mu_\Delta+2\tau\rho) \ .
\end{eqnarray}
Using these relations, 
the structure functions can be  written  in a general
system as functions of $\tau$  
\begin{eqnarray}
w_1 &=& \frac{32}{9}\tau(1+\tau\rho^2)(1+\mu_\Delta+2\tau\rho)G^2
\label{w_1-invariant}\\
w_2 &=& \frac{w_1}{1+\tau\rho^2} = 
\frac{32}{9}\tau(1+\mu_\Delta+2\tau\rho)G^2 \ .
\label{w_2-invariant}
\end{eqnarray}
From the above equations we note that for a pure $M1$ transition
$(1+\tau\rho^2)w_2 -w_1 =0$, hence only the $w_2$ structure
function contributes to the longitudinal response, i.e., 
\begin{equation}
R_L^{M1}(\kappa,\lambda) =
\frac{3{\cal N} \xi_F}{8 m_N\eta_F^3\kappa} 
\frac{\kappa^2}{\tau} 
w_2(\tau) {\cal D}(\kappa,\lambda)
(1-\psi_\Delta^2) \theta(1-\psi_\Delta^2) \ .
\label{RLRFG-magnetic}
\end{equation}

Finally, we end this section by obtaining the relativistic structure 
functions $w_1$
and $w_2$ in the limit $\eta_F=0$. We begin with the expressions
in Eqs.~(\ref{w1d},\ref{w2d}) written in the $\Delta$-system
which, in terms of the dimensionless variables  introduced previously,
read
\begin{eqnarray}
w_1(\tau) &=& \frac12 \mu_\Delta (\mu_\Delta+1)^2
\xi^* \left[G_M^2(\tau)+3G_E^2(\tau)\right]
\label{w1}
\\
w_2(\tau) &=& \frac{1}{2\mu_\Delta} (\mu_\Delta+1)^2
\xi^* \frac{\tau}{\kappa^{*2}}
\left[4 \frac{\tau}{\mu_\Delta^2} G_C^2(\tau)
-G_M^2(\tau)-3G_E^2(\tau)\right]
\ ,
\label{w2}
\end{eqnarray}  
where $\xi^* = (E^*-m_N)/m_N$ is the nucleon kinetic energy in the
$\Delta$ rest frame.
In the $\xi_F=0$ limit the following kinematical relations hold:

\ba
\kappa^{*2} &=& \frac{\tau}{\mu_\Delta^2}(1+\tau\rho^2)
\\
\xi^* &=& \frac{1}{\mu_\Delta} (2\tau\rho+1-\mu_\Delta) \ .
\ea
As a consequence the single-nucleon structure functions reduce to

\ba
w_1(\tau;\xi_F=0) &=& \frac{1}{2} (\mu_\Delta+1)^2
(2\tau\rho+1-\mu_\Delta)
\left(G_M^2+3G_E^2\right)
\\
w_2(\tau;\xi_F=0) &=& \frac{1}{2} (\mu_\Delta+1)^2
\frac{2\tau\rho+1-\mu_\Delta}{1+\tau\rho^2}
\left(4\frac{\tau}{\mu_\Delta^2}G_C^2-G_M^2-3G_E^2\right)
\ea
and the response functions in Eqs.~(\ref{RLxif0},\ref{RTxif0}) read

\ba
R_L^{(0)}(\xi_F=0) &=& \frac{2\tau}{\mu_\Delta^2}(\mu_\Delta+1)^2
(2\tau\rho+1-\mu_\Delta) G_C^2
\\
R_T^{(0)}(\xi_F=0) &=& (\mu_\Delta+1)^2
(2\tau\rho+1-\mu_\Delta)\left(G_M^2+3G_E^2\right) \ .
\ea


\section{Non-relativistic reduction 
          of the delta current}


In this section we present a new ``relativized'' expression for the delta
current that can be implemented very easily in standard non-relativistic
models. Contrary to previous work on the $N\to \Delta$ transition, where an
expansion in the transferred momentum
($q$) and transferred energy ($\omega$) was performed, here we only consider
expansions in powers of the bound nucleon momentum $\eta=p/m_N$, 
keeping the {\em exact} dependence in both $\kappa=q/2m_N$ and
$\lambda=\omega/2m_N$. Thus, we follow closely our previous 
work~\cite{Ama96,Ama98} where new relativized expressions were obtained
for the electroweak nucleon and meson-exchange currents.
It is important to realize that for high-energy conditions
the traditional current operators --- obtained assuming that also $\kappa\ll 1$
and $\lambda\ll 1$ --- are bound to fail, whereas our past work
provides a way to include relativistic aspects into improved, effective
operators for use with the same non-relativistic wave functions.

Following \cite{Jon73} the full $\gamma N \Delta$ vertex
in Eq.~(\ref{JS}) can be decomposed into
magnetic dipole, electric quadrupole
and Coulomb (longitudinal) quadrupole contributions.
In what follows we focus on the contribution of the magnetic term:
due to the smallness of the E2 and C2 form factors, the magnetic term 
is the dominant one for both the
longitudinal and transverse nuclear response functions.
Thus, in the particular case of the
magnetic contribution, the $\gamma N\Delta$ vertex current can be written
in the form \cite{Jon73}
\begin{equation}
\Gamma_{\mu\beta}^M = \frac{G}{m_N^2}
         \epsilon_{\beta\mu}(KQ) \ ,
\label{eq1}
\end{equation}
where we use the notation
$\epsilon_{\beta\mu}(KQ) = \epsilon_{\beta\mu\alpha\gamma}K^\alpha
Q^\gamma$ with $K^\alpha =(P^\alpha +P_\Delta^\alpha)/2$ and $Q^\alpha$
the four-momentum transfer. The form factor $G$ in Eq.~(\ref{G-form-factor})
is proportional to the magnetic form factor $G_M$. 
It can be proven that the relativistic
structure functions obtained with the current in Eq.~(\ref{eq1})
coincide with the general expressions given by
Eqs.~(\ref{w1d},\ref{w2d}) neglecting the electric and Coulomb form factors,
i.e., $G_E=G_C=0$.

In order to proceed with the non-relativistic reduction of the $\gamma
N\rightarrow \Delta$ current, let us rewrite the
transition matrix element for the $\Delta$ excitation
\begin{equation}
\langle \Delta |j_\mu (P_\Delta,P)|N \rangle= 
\overline{u}_\Delta^\beta(P_\Delta,s_\Delta) \Gamma_{\mu\beta}^M u(P,s) \ ,
\label{eq3}
\end{equation}
where $u_\Delta^\beta(P_\Delta,s_\Delta)$ is
the relativistic Rarita-Schwinger spinor describing a spin-3/2 particle
whose general expression is given by \cite{Eri88}
\begin{equation}
u_\Delta^\beta(P_\Delta,s_\Delta) = \sum_{\lambda s'}
{\textstyle \langle \frac12 s' 1 \lambda|\frac32 s_\Delta\rangle}
e_\Delta^\beta (P_\Delta,\lambda) u_\Delta(P_\Delta,s') \ .
\label{eq4}
\end{equation}
Here $e_\Delta^\beta (P_\Delta,\lambda)$ are the spherical vectors boosted
to momentum $\np_\Delta$ and $u_\Delta(P_\Delta,s')$ are spin-$1/2$ Dirac
spinors.
Introducing the dimensionless variables
$\neta_\Delta=\frac{\np_\Delta}{m_\Delta}$ and
$\varepsilon_\Delta= \frac{E_\Delta}{m_\Delta}$, we can simply write
\begin{eqnarray}
e^\beta_\Delta(P_\Delta,\lambda) 
&=&
\left( \nne_\lambda\cdot\neta_\Delta,\,
       \nne_\lambda+
       \frac{\nne_\lambda\cdot\neta_\Delta}{1+\varepsilon_\Delta}
       \neta_\Delta
\right)
\label{eq5}\\
u_\Delta(P_\Delta,s') 
&=& \sqrt{\frac{1+\varepsilon_\Delta}{2}}
    \left[
    \begin{array}{c}
      1 \\
      \frac{\nsigma\cdot\neta_\Delta}{1+\varepsilon_\Delta}
     \end{array}
     \right]\chi_{s'} \ .
\label{eq6}
\end{eqnarray}
Thus, following the same procedure developed in \cite{Ama96,Ama98},
the matrix element for a general four-dimensional
gamma matrix $\Gamma$ can be related to the corresponding bi-spinor matrix
element according to
\begin{equation}
\overline{u}_\Delta\Gamma u 
\equiv \chi^{\dagger}_{s'}\overline{\Gamma}\chi_s
=\chi^{\dagger}_{s'}
\frac{1}{\sqrt{2(1+\varepsilon_\Delta)}}
\left[(1+\varepsilon_\Delta)\Gamma_{11}
     -\nsigma\cdot\neta_\Delta
      \Gamma_{21}
\right]
\chi_{s}
+O(\eta) \ ,
\label{eq7}
\end{equation}
where the expansion only to leading order in $\eta$ has been considered.

To make the discussion that follows easier we introduce new
dimensionless variables which are analogous to the ones already used
in the case of the nucleon quasi-elastic peak \cite{Ama96,Ama98}. In
particular, let us define the variable $\lambda'$ given by the relation:
$\varepsilon_\Delta=1+2\lambda'$. 
It is straightforward to
derive the following relations (valid up to first order in $\eta$)

\begin{eqnarray}
\lambda'
&=& \frac{1}{\mu_\Delta}\left(\lambda-\frac{\mu_\Delta-1}{2}\right)
=\tau'+\neta\cdot\nkappa'+O(\eta^2)
\label{eq8}\\
\tau'&=&\frac{1}{\mu_\Delta}
      \left[\tau
          + \left(\frac{\mu_\Delta-1}{2}\right)^2
       \right] 
\label{eq9}\\
\nkappa'&=&\frac{\nkappa}{\mu_\Delta};
\kern 0.7cm
\neta_\Delta=2\nkappa'+\frac{\neta}{\mu_\Delta}
\label{eq10}
\end{eqnarray}
with $\mu_\Delta$ as defined in Eq.~(\ref{VAR}).
The current operator $\overline{\Gamma}$ in Eq.~(\ref{eq7}) can be then
written up to leading order in $\eta$ in the form
\begin{equation}
\overline{\Gamma} =
\frac{1}{\sqrt{1+\tau'}}
\left[
       (1+\tau')\Gamma_{11}
      -\nsigma\cdot\nkappa'\Gamma_{21}
\right]
+O(\eta) \ .
\label{eq11}
\end{equation}
Moreover, in terms of the spin transition operator for the $\Delta$
in Eq.~(\ref{SPIN}), the general matrix elements between 1/2 and 3/2 spinors
result
\begin{eqnarray}
\overline{u}^0_\Delta\Gamma u
&=& {\textstyle 
        \langle \frac32 s_\Delta|
        2{\nS}^{\dagger}\cdot\nkappa'
        \overline\Gamma
        |\frac12 s\rangle
     }
+O(\eta)
\label{eq12}\\
\overline{u}^i_\Delta\Gamma u
&=& {\textstyle 
        \langle \frac32 s_\Delta|
    }
        \left[{\nS}^{\dagger}+
               \frac{{\nS}^{\dagger}\cdot\nkappa'}{1+\tau'}
               2\nkappa'
        \right]^i 
        \overline\Gamma
 {\textstyle 
        |\frac12 s\rangle
 } +O(\eta) \ .
\label{eq13}
\end{eqnarray}

The non-relativistic reduction (valid up to leading order in $\eta$)
for the magnetic $N\Delta$ current operator,
$J_\mu$, is defined through the relation
\begin{equation}
{\textstyle \langle \frac{3}{2}s_\Delta |J_\mu|\frac12 s\rangle}=
\langle \Delta |j_\mu (P_\Delta, P)|N\rangle 
\end{equation}
and can be obtained by using
the general relations given by Eqs.~(\ref{eq8}-\ref{eq10}) and taking into
account the fact that the magnetic current operator 
$\Gamma_{\mu\beta}^M$ is
diagonal in spin space. The time and space components are given by
\begin{eqnarray}
{J}_0 &\simeq&
\frac{G}{m_N^2}\sqrt{1+\tau'}
        \left[{\nS}^{\dagger}+
               \frac{{\nS}^{\dagger}\cdot\nkappa'}{1+\tau'}
               2\nkappa'
        \right]^i 
\epsilon_{i0}(KQ) 
\label{eq14}\\
{J}_i &\simeq&
\frac{G}{m_N^2}
 \sqrt{1+\tau'}\left\{2{\nS}^{\dagger}\cdot\nkappa'\epsilon_{0i}(KQ)
+
         \left[{\nS}^{\dagger}+
               \frac{{\nS}^{\dagger}\cdot\nkappa'}{1+\tau'}2\nkappa'
         \right]^j
         \epsilon_{j i}(KQ)\right\} \ .
\label{eq15}
\end{eqnarray}
Using the relations
\begin{eqnarray}
\epsilon_{i0}(KQ)&=&
\epsilon_{i0jk}K^jQ^k
\simeq -2m_N^2(\neta\times\nkappa)_i
\label{eq16}\\
\epsilon_{ji}(KQ)&=&
\epsilon_{ji\alpha\beta}K^\alpha Q^\beta \simeq
2m_N^2\epsilon_{jik}\kappa^k
\label{eq17}
\end{eqnarray}
and taking into account the fact that $\nkappa'$ and $\nkappa$ are 
parallel vectors 
(see Eq.~(\ref{eq10})), we finally get
\begin{eqnarray}
{J}_0 &=&
-2 G\sqrt{1+\tau'}{\nS}^{\dagger}\cdot(\neta\times\nkappa) +O(\eta^2)
\label{J0-nr}\\
{\nJ}
&=& 2 G \sqrt{1+\tau'}
    (\nS^{\dagger}\times\nkappa) + O(\eta) \ .
\label{eq19}
\end{eqnarray}
Note that the  current ${\nJ}$ is of order $O(1)$,
whereas the charge ${J}_0$ 
is of order $O(\eta)$. Thus, one expects the
contribution of the $N\Delta$ current to be considerably more
important for the transverse response. 

It is important to note that in order to be consistent one should
treat the charge and current at the same level and perform an
expansion of the current to order $O(\eta)$. Such program can be
carried out by using the techniques developed in 
\cite{Ama96,Ama98}. However, for the present case an additional
simplification can be made by using the expression of the
charge operator to order $O(\eta)$ in Eq.~(\ref{J0-nr}), and taking into
account the fact that, before performing the expansion, the original 
current was gauge-invariant.
This invariance property should be valid for all the orders in the
expansion (we are not expanding in $\nq$ or $\omega$).
Hence we can relate the longitudinal component
of the current with the density, i.e.,
\begin{equation}
\nJ\cdot\nkappa = \lambda J_0 
= - 2G\lambda \sqrt{1+\tau'}(\nS^{\dagger}\times\neta)\cdot\nkappa \ .
\end{equation}
Hence we can obtain an improved and gauge-invariant
current by adding to the expression in Eq.~(\ref{eq19})
a new piece of order $O(\eta)$ given by
\[\nJ_{\rm gauge}= -2G\lambda\sqrt{1+\tau'}(\nS^{\dagger}\times\neta)
\ .\]

Of course there could be
additional corrections of order $O(\eta)$ in the transverse current
that cannot be fixed
by just using the continuity equation. However, as we will show
below, the transverse response function computed with this improved
current differs from the exact relativistic results by 
terms only of order $O(\eta^2)$, proving the high quality of this expansion
for most applications in nuclear physics. 
Therefore the new expression of the current that we will consider below is
\begin{eqnarray}\label{J-final}
\nJ &=& 2G\sqrt{1+\tau'}\,\nS^{\dagger}\times(\nkappa-\lambda\neta) \ .
\end{eqnarray}


In order to test the quality of the above expansion of the current 
we proceed to compute the response functions in a non-relativistic
Fermi gas using relativistic kinematics and the new currents
in Eqs.~(\ref{J0-nr},\ref{J-final}). We begin by computing the analytical
expressions and compare with the exact relativistic answer.
Results are shown and discussed in the next section.

The calculation of the non-relativistic nucleon tensor
needed to evaluate the nuclear response functions can be done
by performing the following traces
\begin{eqnarray}
f^{nr}_{00} 
& = & \frac43 \frac{m_\Delta}{m_N}\,{\rm Tr}\,[{J}^{\dagger}_0 {J}_0]
\equiv w^L_{nr}\neta_T^2
\label{f00-nr}\\
f^{nr}_{11}+f^{nr}_{22}
& = & \frac43\frac{m_\Delta}{m_N}
\,{\rm Tr}\, [{\nJ}_T^{\dagger}\cdot{\nJ}_T]
\equiv  2w^T_{nr} \ ,
\label{eq21}
\end{eqnarray}
where we have introduced the longitudinal and transverse,
non-relativistic structure functions
$w^L_{nr}$ and $w^T_{nr}$
\begin{eqnarray}
w^L_{nr} &=&\frac43\frac{16m_\Delta }{3m_N} G^2(1+\tau')\kappa^2 
\label{wL-nr}\\
w^T_{nr} &=& \frac43\frac{16m_\Delta }{3m_N}  G^2(1+\tau') 
(\kappa^2-2\lambda\nkappa\cdot\neta)+O(\eta^2) \ .
\end{eqnarray}

In order to compare the relativistic
and non-relativistic response functions it is convenient to use the
kinematical relations:
\begin{eqnarray}
1+\tau' &=& \frac{1}{2\mu_\Delta}(1+\mu_\Delta+2\tau\rho)
\\
\kappa^2-2\lambda\nkappa\cdot\neta &=& \tau(1+\tau\rho^2)+O(\eta^2) \ .
\end{eqnarray}
We can then write, up to first order, the relations
\begin{eqnarray}
w^T_{nr} &=& w_1+ O(\eta^2) \\
w^L_{nr} &=& \frac{\kappa^2}{\tau}\frac{w_1}{1+\tau\rho^2}+O(\eta^2) 
            = \frac{\kappa^2}{\tau}w_2 + O(\eta^2) \ , 
\end{eqnarray}
where $w_1$ and $w_2$ are 
the magnetic relativistic functions given
in Eqs.~(\ref{w_1-invariant},\ref{w_2-invariant}).

The nuclear response functions are given finally by
\begin{eqnarray}
R_L^{nr}
&=& \frac{3{\cal N}\xi_F}{8m_N\eta_F^3\kappa}
     \theta(1-\psi_\Delta^2)(1-\psi_\Delta^2)
     w^L_{nr}{\cal D}
\label{RLnr}\\
&=& \frac{3{\cal N}\xi_F}{8m_N\eta_F^3\kappa}
    \theta(1-\psi_\Delta^2)(1-\psi_\Delta^2)
    \frac{\kappa^2}{\tau}\left[w_2+O(\eta^2)\right]{\cal D}
\nonumber\\
R_T^{nr}
&=& \frac{3{\cal N}\xi_F}{8m_N\eta_F^3\kappa}
    \theta(1-\psi_\Delta^2)
    (1-\psi_\Delta^2) 
    2w^T_{nr}
\label{RTnr}
\\
&=& \frac{3{\cal N}\xi_F}{8m_N\eta_F^3\kappa}
     \theta(1-\psi_\Delta^2)
    (1-\psi_\Delta^2) 
     2\left[w_1 +O(\eta^2)\right]
\nonumber
\end{eqnarray}
with ${\cal D}$ as given by Eq.~(\ref{Delta}).
Although we call these functions ``non-relativistic'', actually they
contain enough relativistic ingredients to be high-quality
approximations to the exact RFG result. In fact, first we
use relativistic kinematics, so that the phase space and momentum
integrals are done exactly. Second, we consider the new currents
expanded to include effects up to order $O(\eta)$, so that the 
dynamics of the problem are
correct to that order. Comparing these expressions with
the exact relativistic responses in Eqs.~(\ref{RTRFG},\ref{RLRFG-magnetic}),
we see that the relative differences between 
$R_L^{nr}$ and $R_L$, and between   
$R_T^{nr}$ and $R_T$ 
are of order $O(\eta^2)$.
In the next section we test numerically the quality of the non-relativistic 
responses of Eqs.~(\ref{RLnr},\ref{RTnr}).


\section{Results}


In this section we present numerical results for the relativistic response
functions in the region of the $\Delta$-peak and check numerically the
 quality of
our new approximation to the $N\to\Delta$ electromagnetic current.
We also investigate the importance of different contributions in the
relativistic responses. We shall present results for medium and high 
momentum transfers, ranging from $q=0.5$  to 2 GeV/c, and thus 
will also be able to analyze the validity of the traditional non-relativistic
calculations performed for different values of $q$. 

Before starting our analysis it is convenient to check our model
with some of the available experimental data \cite{Bat72}. In this way we 
can fix some of the ingredients that enter in the
model, in particular, the magnetic form factor $G_M$ of the $\Delta$ and
the modification of the response due to the finite
$\Delta$ width as a consequence of its later decay 
into the $N$-$\pi$ channel, which we incorporate by performing
a convolution with the responses for stable particles. 

In fig.~1 we show the transverse cross section $\sigma_T$ for
the inclusive reaction H$(e,e')$ from the nucleon. 
The transverse nuclear cross section is defined by 
\begin{equation}
\sigma_T = \hbar c \frac{2\pi\alpha^2}{\omega+\frac{Q^2}{2m_N}}r_T \ ,
\end{equation}
where $r_T=\frac{m_\Delta}{E_\Delta}2w_1$ is the transverse response of a
single nucleon. 
In this calculation we have included the $\Delta$ width by
substituting for the energy-conserving delta function a Lorentzian shape
\begin{equation}
\delta(\omega+m_N-E_\Delta)
\longrightarrow 
\frac{E_\Delta}{m_\Delta}
\frac{1}{\pi}
\frac{\Gamma(s)/2}{(\sqrt{s}-m_\Delta)^2+\Gamma(s)^2/4} \ ,
\end{equation}
where $s=(m_N+\omega)^2-q^2$ is the invariant mass of the 
initial photon and nucleon. Results that
include a width $\Gamma(s)$ (which is zero at threshold
and equal to the width $\Gamma_0= 120$ MeV at resonance)
are represented by solid lines in fig.~1.
The dependence of $\Gamma(s)$
on the invariant mass is given by \cite{Gil97}
\begin{equation}
\Gamma(s) = \Gamma_0 
            \frac{m_\Delta}{\sqrt{s}}
            \left(\frac{p_\pi^*}{p_\pi^{res}}\right)^3 \ ,
\end{equation}
where $p_\pi^*$ is the momentum of the final pion resulting from the 
$\Delta$ decay (in the $\Delta$-system) given by
\begin{equation}
p_\pi^* = \frac{1}{\sqrt{s}}
\left[ \frac{(s-m_N^2-m_\pi^2)^2}{4}-m_N^2m_\pi^2\right]^{1/2} 
\end{equation}
and $p_\pi^{res}$ is its value at resonance, obtained from the above
expression for 
$\sqrt{s}=m_\Delta$. 
For comparison, we show in fig.~1 with dashed lines results obtained 
by considering a constant width $\Gamma=\Gamma_0$.

Other ingredients that enter in the cross section are the $\Delta$
form factors. We use the parameterization
\begin{eqnarray}
G_M(Q^2) &=& G_M(0)f(Q^2)\\
G_E(Q^2) &=& G_E(0)f(Q^2) \ ,
\end{eqnarray}
where we assume that the same dependence in $Q^2$ is valid for the
electric and magnetic form factors, given by the function \cite{Weh89}
\begin{equation}
f(Q^2)= G_E^P(Q^2)
\left(1-\frac{Q^2}{3.5\,{\rm (GeV/c)}^2}\right)^{-1/2}
\end{equation}
with $G_E^P$ the electric form factor of the proton, for which we use
the Galster parameterization $(1+4.97\tau)^{-2}$ \cite{Gal71}.
The above equation reflects the fact that the isobar form factor 
falls off faster than the proton form factor.
Unless otherwise indicated, we take $G_C=0$ and use the following values 
\cite{Jon73} of the form factors at the origin 
\begin{equation}\label{form-factors}
G_M(0)=2.97,  \kern 2cm G_E(0)=-0.03.
\end{equation}
Later on we show the effect on the longitudinal response function 
introduced by considering a C2 form factor that is different from
zero.

Our calculation (solid lines) displayed in fig.~1 is slightly below
the data for the three values of the momentum transfer 
$Q^2=-0.2,-0.3,-0.4$ (GeV/c)$^2$, reflecting the fact that we have not
included the background contributions of  
non-resonant pion production, which
produce an additional increase of the cross section. 

In fig.~2 we show results for the nuclear inclusive cross section 
per nucleon from $^{12}$C compared with the experimental data taken from
\cite{Ang96,Bar83}. Dotted lines correspond to the RFG 
for a stable $\Delta$. 
A more realistic model of the $\Delta$ peak requires the inclusion of the 
$\Delta$ width in the cross section, which we show with dashed lines.
The nuclear responses including the width, $R_\Gamma(q,\omega)$, 
are computed from the responses $R(q,\omega,W)$ 
for a stable $\Delta$ with mass $W$  by a convolution
\begin{equation}
R_\Gamma(q,\omega)= \int_{m_N+m_\pi}^{W_{max}} 
\frac{1}{\pi}\frac{\Gamma(W)/2}{(W-m_\Delta)^2+\Gamma(W)^2/4}
R(q,\omega,W)dW \ ,
\end{equation}
where the integration interval goes from threshold
to the maximum value allowed in the Fermi gas model, 
$W_{max}^2 = (E_F+\omega)^2-(q-k_F)^2$. 
The inclusion of the $\Delta$ width produces a broadening  of the
$\Delta$ peak and correspondingly a decrease of the strength. 

As an illustration of how one could improve the model in the 
quasi-elastic-peak region, we also show with dot-dashed lines the 
quasi-elastic cross section computed with the PWIA model of ref.
\cite{Ama96b}. Here the mean difference with the RFG model is the
inclusion of the momentum distribution of the finite-sized nucleus,
which produces the ``tails'' of the cross section, and the 
binding energy of the nucleons in the nucleus, which produces a shift to higher
energies (in the direction of data). 
Here for the PWIA calculation we use relativistic kinematics,
final states are described as plane waves and the
electromagnetic current used contains relativistic corrections 
to order $\eta$. 
Finally, we show with solid lines the results computed with a hybrid 
model in which we add the PWIA cross section for the quasi-elastic 
contribution to the RFG result for the $\Delta$ contribution.

As we can see in fig.~2, our results are  below the data 
in the dip and $\Delta$ region. This was expected because other
contributions coming mainly from two-nucleon emission and 
non-resonant pion production (not included in our model)
also enter here \cite{Van81,Gil97,Dek91,Ama94}. 
However, our intention in this work is not to reproduce the
experimental data nor to present a complete
model including all of the physical contributions in this energy region,
but instead to discuss the effect of different ingredients
in the calculation and present a new set of improved currents specifically
for excitation of the $\Delta$ peak that now include the relevant
relativistic content --- these could now straightforwardly be used in 
standard non-relativistic many-body models 
with relativistic kinematics. 

The quality of the new approximation to the relativistic, magnetic
$\Delta$ current is shown in fig.~3, where the exact RFG longitudinal
and transverse responses using magnetic and electric form factors 
are displayed with solid lines. 
Here we show just the $\Delta$ contribution to the responses.
In addition we show with dashed lines the responses computed in
the non-relativistic Fermi gas model with relativistic kinematics
and the new currents in Eqs.~(\ref{J0-nr},\ref{J-final}).
For comparison we also show with dot-dashed lines results for
the non-relativistic Fermi gas model and relativistic kinematics, but using the 
traditional non-relativistic current. The improvement of the
description of the relativistic results using our currents is clear
from this figure --- the solid and dashed lines almost coincide.
This proves that our expansion to order $O(\eta)$ is precise enough 
to describe the $\Delta$ excitation in nuclei with negligible error 
for high momentum transfers.

In fig.~4 several relativistic effects and ingredients of the
 calculation are analyzed.  
Therein we show the longitudinal and transverse responses for 
$q=0.5,1$ and 2 GeV/c. With solid lines we show the $\Delta$ peak 
computed within our model, while  with dashed lines we show the 
$\Delta$ peak computed using the Peccei Lagrangian \cite{Pec69}.
The Peccei Lagrangian only includes the first coupling
 $\Gamma^1_{\mu\nu}$ given in Eq.~(\ref{Gamma-1}) with coupling
 constant $C_1(0)= 2.5$ GeV$^{-1}$. This value has been chosen so that
the transverse response is equal to the one computed with the full
Lagrangian in the $\Delta$ peak 
for $q=500$ MeV/c.  The importance of using the full
vertex in Eq.~(\ref{Gamma}) is clear from this figure. First, although the 
coupling constants can be chosen so that the 
transverse responses computed with both Lagrangians are
similar for moderate $q=500$ MeV/c, they begin to fail for higher 
$q$-values. These differences are seen to be most important in
the longitudinal
response, where the Peccei Lagrangian clearly gives an extremely
large result. The reason for this unphysical behavior of the Peccei
Lagrangian 
is that in Eq.~(\ref{fst00})
there are important cancelations among the $C_1$, $C_2$ and $C_3$ 
pieces in the longitudinal channel \cite{Duf68}.
Second, as the Peccei Lagrangian is the one usually employed to compute 
the MEC contribution involving virtual $\Delta$ excitation \cite{Van81}, 
it is mandatory to use the full Lagrangian --- or at least the magnetic
piece --- if one wants to compute the
longitudinal contribution of the MEC in this channel. 

As reference, 
in fig.~4 we also show with dot-dashed lines the quasi-elastic peak
responses. For high $q$-values the two peaks overlap and the
importance of the $\Delta$ in both responses also increases
with $q$. 
This is better seen in fig.~5 where we 
also show the effect on the response functions
produced by incorporating the finite $\Delta$ width (solid lines).
Dashed lines correspond to results without including the
isobar finite width, while dot-dashed lines correspond to the 
quasi-elastic peak. 
In the above results no Coulomb form factor has been included.
In this case
the contribution of the $\Delta$ in the longitudinal 
response is due to the Fermi motion of the nucleons inside the
nucleus \cite{Cha93}, the main contribution here coming from the magnetic 
$\Delta$ excitation, which is zero only for nucleons at rest.
This is better seen in Eq.~(\ref{f00-nr}) where the longitudinal,
magnetic single-nucleon response is seen to be proportional to
$\eta_T^2$ and to the function $w_L^{nr}$ which is proportional to
$\kappa^2$ (see Eq.~(\ref{wL-nr})), explaining the
increase with $q$ of the longitudinal response observed in fig.~5.
In the static limit, the longitudinal response becomes proportional to
the Coulomb $C2$ form factor, as shown at the end of sect.~2.

In fig.~6 we show the dependence of the longitudinal response on the
Coulomb form factor of the $\Delta$. With solid lines we show the 
$\Delta$ peak without C2 multipoles, while the dashed lines include a
Coulomb contribution with form factor
\[ G_C(Q^2)= -0.15 G_M(Q^2)\ . \]
We can see that the longitudinal response is quite sensitive to this form
factor, especially for high $q$.
This can be easily understood from the analytical expression of the 
structure functions in Eqs.~(\ref{w1d},\ref{w2d}). Only the structure
function $w_2$ depends on $G_C$ and its dependence on the form factors
is carried by the quadratic combination 
$\frac{Q^2}{m_\Delta^2}G_C^2-G_M^2-3G_E^2$, which is not very sensitive
to small values of $G_C$. However looking at $R_L$ in Eq.~(\ref{RLRFG}),
we see that the correction due to the  term $w_2{\cal D}$ is small;
hence the main contribution comes from the 
combination $(1+\tau\rho^2)w_2-w_1$. 
The important point is that, if
no C2 term is present, that combination is exactly zero, i.e.,
$(1+\tau\rho^2)w_2-w_1=0$. Thus, the only contribution to the 
longitudinal response comes from the higher-order term
$w_2{\cal D}$. This explains why the longitudinal isobar 
response is so small. On the other hand, if the
C2 form factor is nonzero,
there will be a contribution from the leading-order term 
$(1+\tau\rho^2)w_2-w_1$ which is proportional to $G_C^2$. This term,
although small, is of the same order of magnitude as the 
$w_2{\cal D}$ piece. 
Therefore, we can conclude (at least for values of $G_C$ such as those
assumed here) that a correct treatment
of the isobar longitudinal response requires the inclusion of the 
Coulomb form factor. 

To finish this section, we have also explored the sensitivity
of the response functions to inclusion of the electric E2 form factor,
finding that both responses are quite insensitive. 
The reason is that the E2 contribution always adds incoherently 
to the M1 form factor in the combination $G_M^2+3G_E^2$, as seen in
Eqs.~(\ref{w1d},\ref{w2d}). Therefore, the inclusion of a small E2
form factor does not significantly modify any of the responses. 
In order to obtain appreciable effects due to the electric 
form factor, one could explore other observables where 
interferences can occur; for instance, one
could analyze the angular distribution of the 
$\Delta$ emission in the transverse channel \cite{Noz90}.


\section{Conclusions}


In summary, in this paper we have obtained a new expansion of 
the relativistic $\Delta$ electro-excitation current 
to first order in $\eta=p/m_N$, maintaining the exact dependence on
the momentum and energy transfers.  We have tested this
new expansion  by performing a calculation using a non-relativistic
Fermi gas model together with relativistic kinematics. The resulting
longitudinal and transverse responses are found to be very close to the
exact result computed with the RFG model.
Therefore it is expected that the use of this current 
will provide a significant improvement when used in more realistic
models of inclusive electron scattering from nuclei.  

We have also performed a comparison between different Lagrangians in
the treatment of the $\Delta$ excitation, finding that the Peccei
Lagrangian is inappropriate in the longitudinal channel 
for all of the  $q$-values analyzed. 
We have studied the contribution of the isobar emission to the 
longitudinal response, in general finding a small contribution  
for medium $q$-values, although increasing significance as $q$
increases. In our calculations we include  
the finite width of the $\Delta$ assuming  a Lorentzian shape.
With regards to the quadrupole amplitudes of the $\Delta$,
 we have found a large sensitivity of the longitudinal
response to inclusion of the Coulomb form factor of the isobar, 
especially for high
$q$, a fact that could be of importance for the present investigations
of the longitudinal nuclear response. 
On the other hand, both L and T responses are found to be insensitive to the 
quadrupole E2 form factor.

Finally, it is of interest to extend the present studies to 
the two-particle emission channels which provide important contributions 
in this energy region. In particular, one expects these ideas to be 
relevant in the analysis of non-resonant 
pion production $(e,e'p\pi)$ and two-nucleon emission $(e,e'2N)$ reactions,
where there is the
hope of extracting detailed information on short-range
correlations in nuclei. It is clear that, in order 
to extract ground-state properties from these reactions,  
high-$q$ values are needed, since one wishes to reach a reasonably 
quasi-free regime in which final-state interaction effects are expected
to be minimal. However, for such kinematical conditions, as our
present studies indicate, relativistic effects
play a role and it is important to have them theoretically under control. 
At this point in time we have explored ways to relativize the
electroweak currents for selected situations --- our intent is to
continue along the same path for other related problems such as those
mentioned above.


\section*{Acknowledgments}


J.E.A. wants to thank J. Nieves  for helpful discussions on the $\Delta$ 
electro-excitation.

This work is supported  in part by funds provided
by the U.S.  Department of Energy (D.O.E.) under cooperative agreement
\#DE-FC01-94ER40818, in part  by DGICYT (Spain) under Contract Nos.
PB95-1204, PB95-0123 and PB95-0533-A and the Junta de Andaluc\'{\i}a
(Spain), in part by NATO Collaborative Research Grant \#940183 and
in part by the Spanish-Italian Research Agreement HI1998-0241.


\newpage 
{\bf Figures}

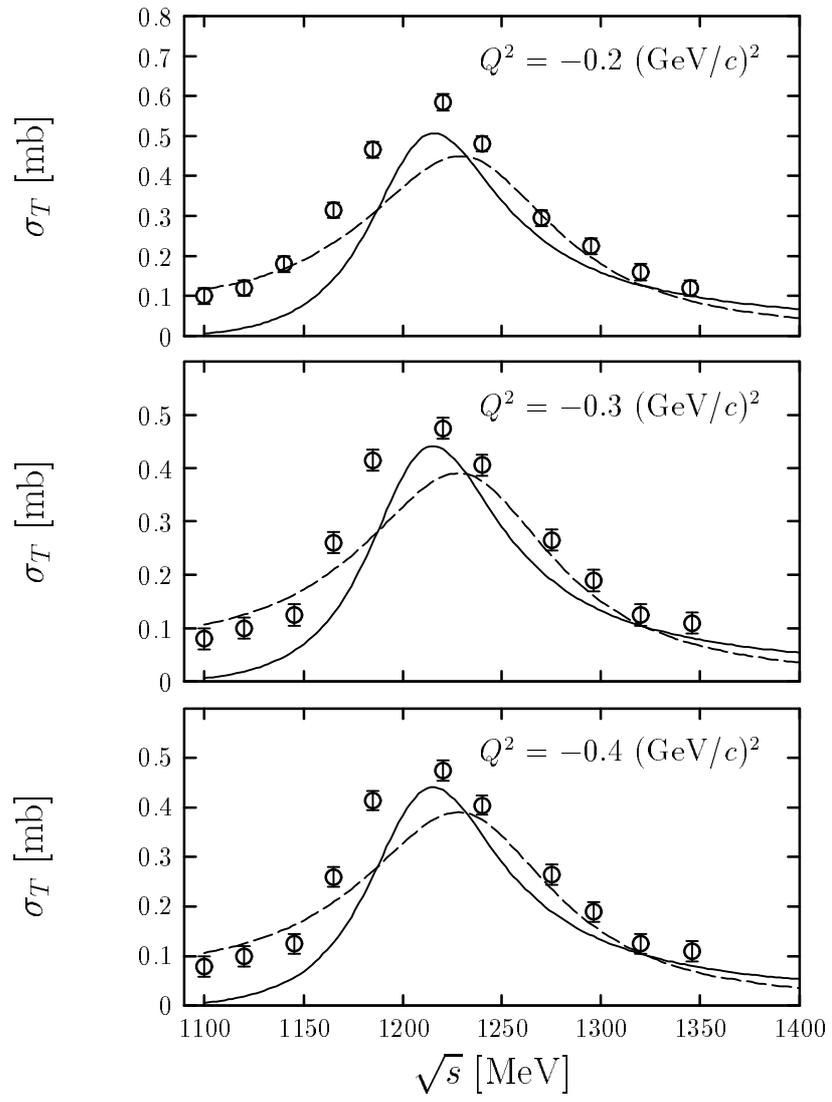
{\bf Figure 1:} Transverse cross section $\sigma_T$ 
for the inclusive reaction H$(e,e')$ in the $\Delta$ peak 
as a function of the $\Delta$ invariant mass $\sqrt{s}$.
The solid lines include a width $\Gamma(s)$ which is zero at threshold.
The dashed lines use a constant width $\Gamma=120$ MeV.
Experimental data are from \cite{Bat72}.

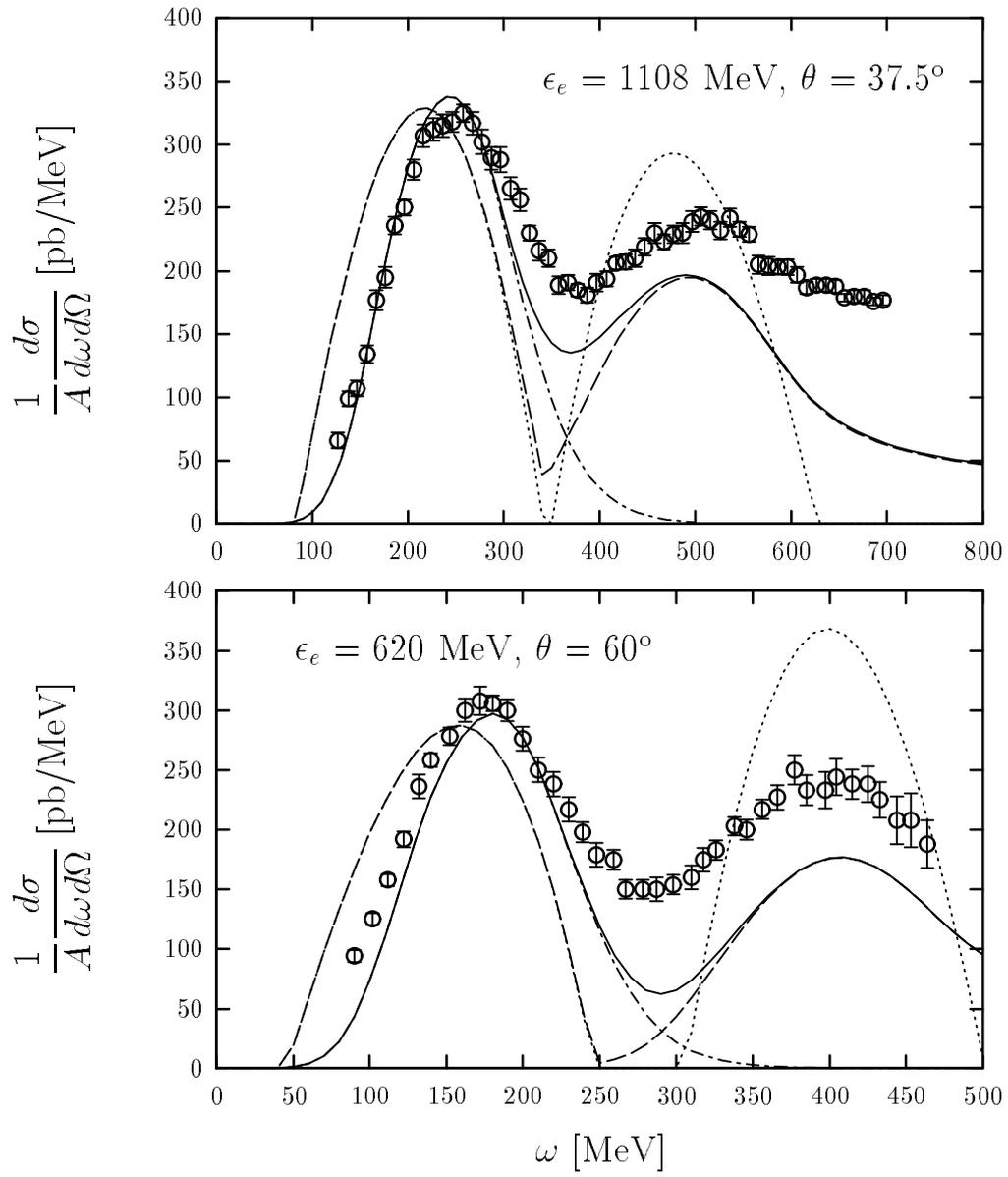
{\bf Figure 2:} Inclusive cross section per nucleon
 from $^{12}$C, for two beam energies and two scattering angles.
Dotted lines: RFG without $\Delta$ width; dashed: RFG including the
finite  $\Delta$ width; dot-dashed: PWIA for the quasi-elastic peak;
solid: hybrid model obtained by adding the PWIA cross section 
in the quasi-elastic 
peak and RFG cross section in the $\Delta$-peak. 
Experimental data are from \cite{Ang96,Bar83}.

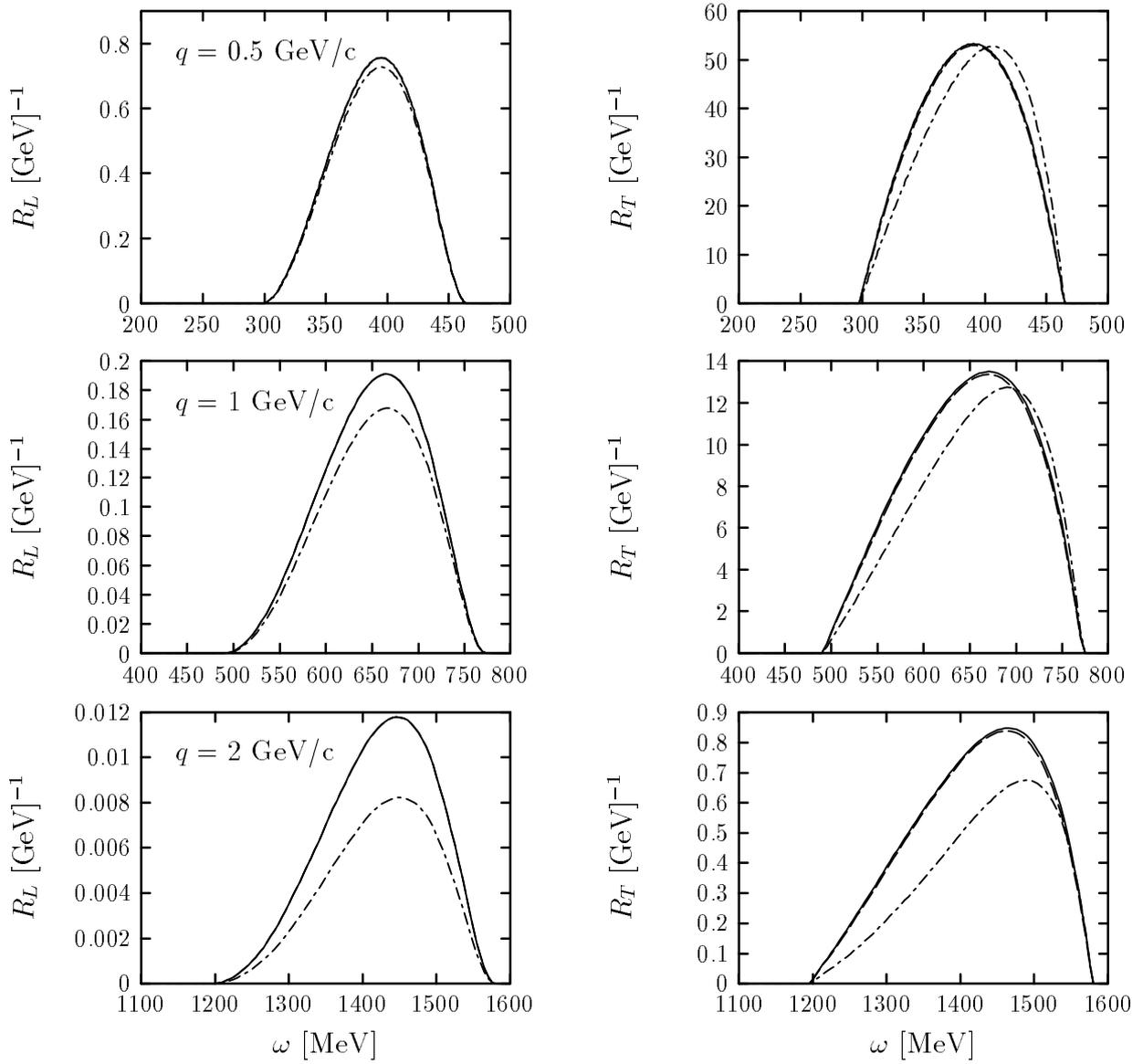
{\bf Figure 3:} Electromagnetic responses in the $\Delta$-peak
for $^{12}$C with $k_F=225$ MeV/c, without a $\Delta$-width.
 Solid: exact results within the RFG model;
dashed: new expansion of the electromagnetic current to 
first  order in $\eta$;
dot-dashed: traditional non-relativistic current. 
We use relativistic kinematics in all cases.

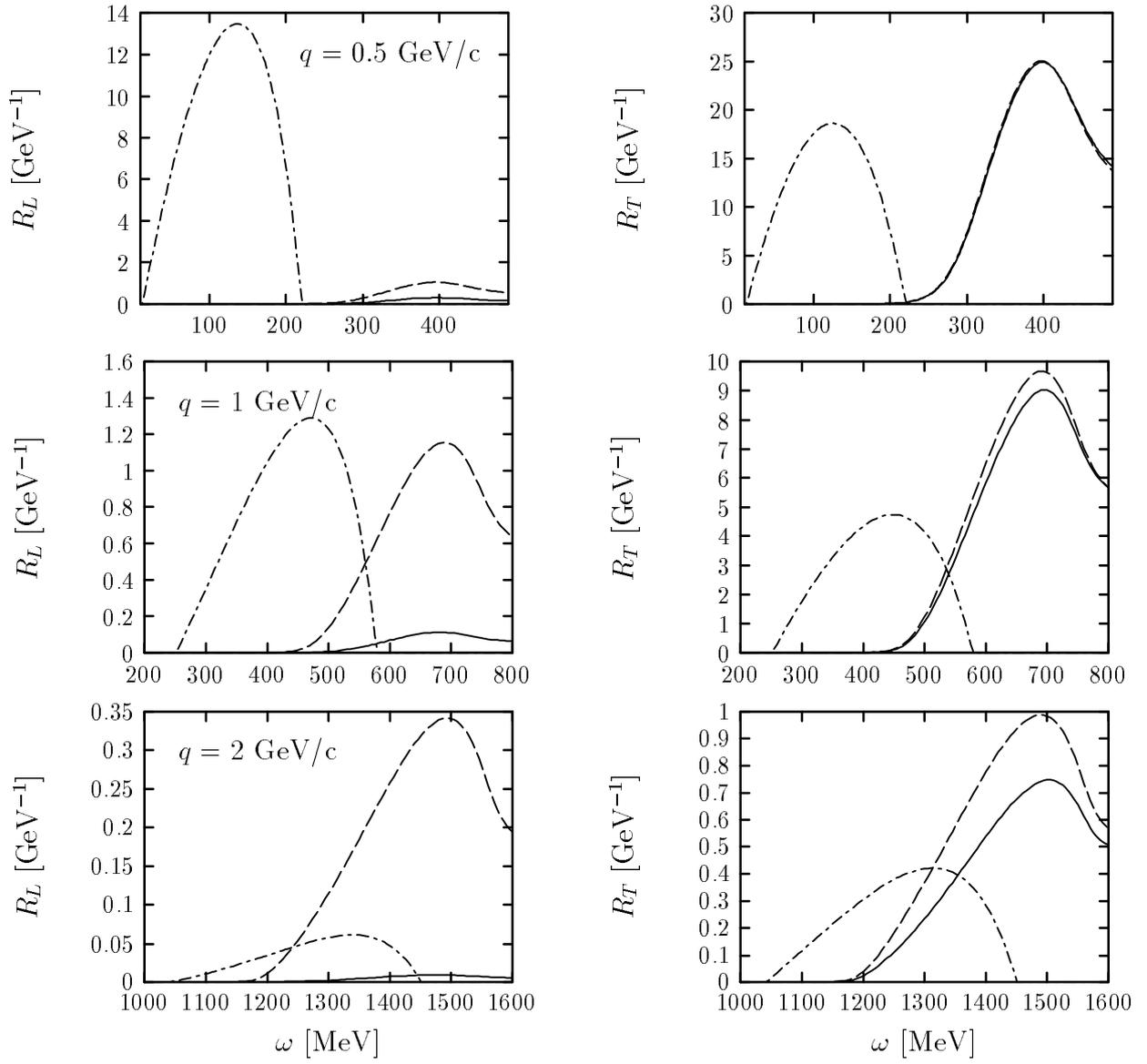
{\bf Figure 4:} Electromagnetic responses 
for $^{12}$C in the RFG model with $k_F=225$ MeV/c, with $\Delta$-width.
 Solid: $\Delta$-peak Using the Jones \& Scadron Lagrangian;
dashed: $\Delta$-peak using the Peccei Lagrangian;
dot-dashed: quasi-elastic peak.

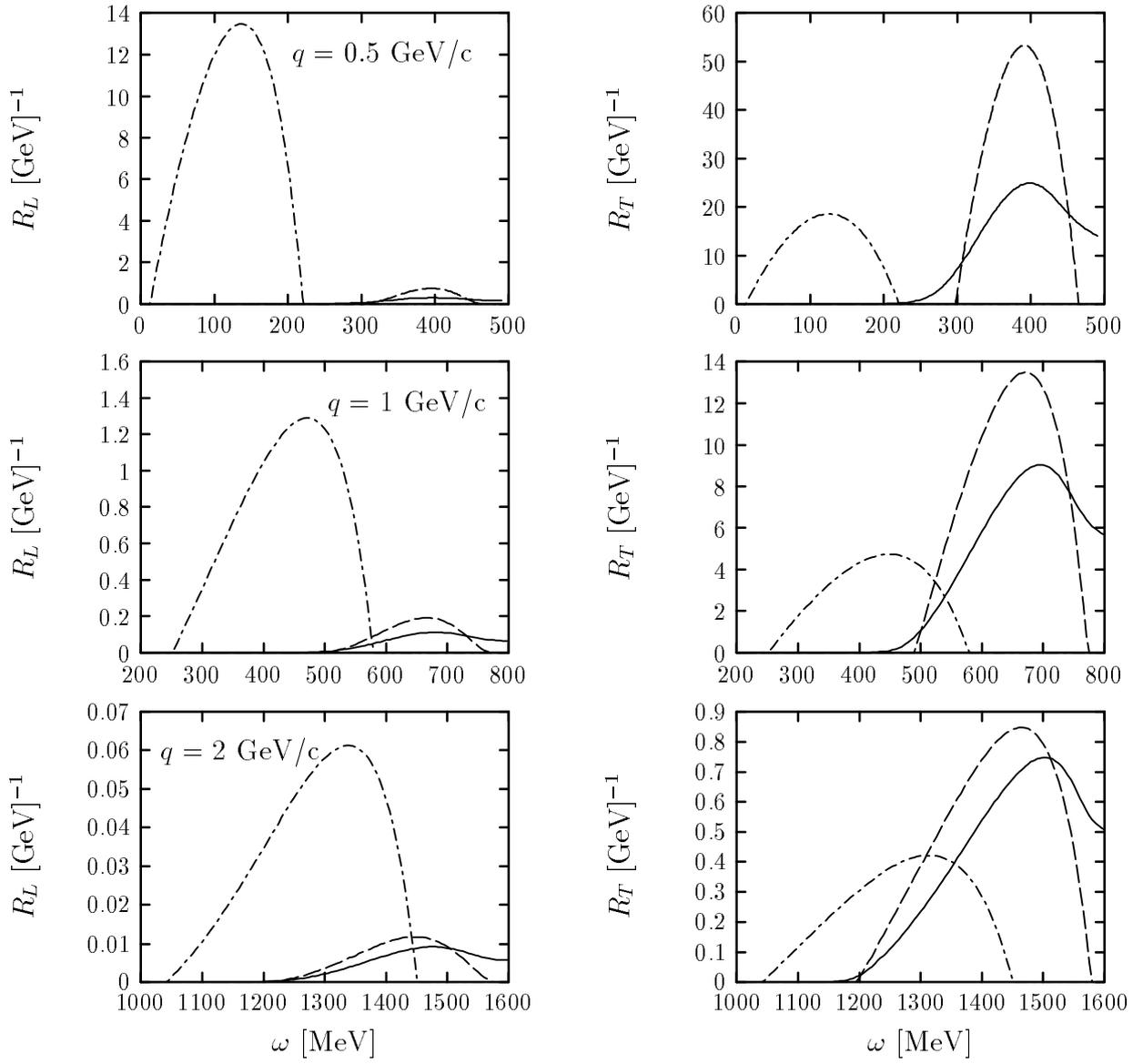
{\bf Figure 5:} Electromagnetic responses 
for $^{12}$C in the RFG model with $k_F=225$ MeV/c.
 Solid: $\Delta$-peak  including the  $\Delta$-width;
dashed: $\Delta$-peak for stable isobar;
dot-dashed: quasi-elastic peak.

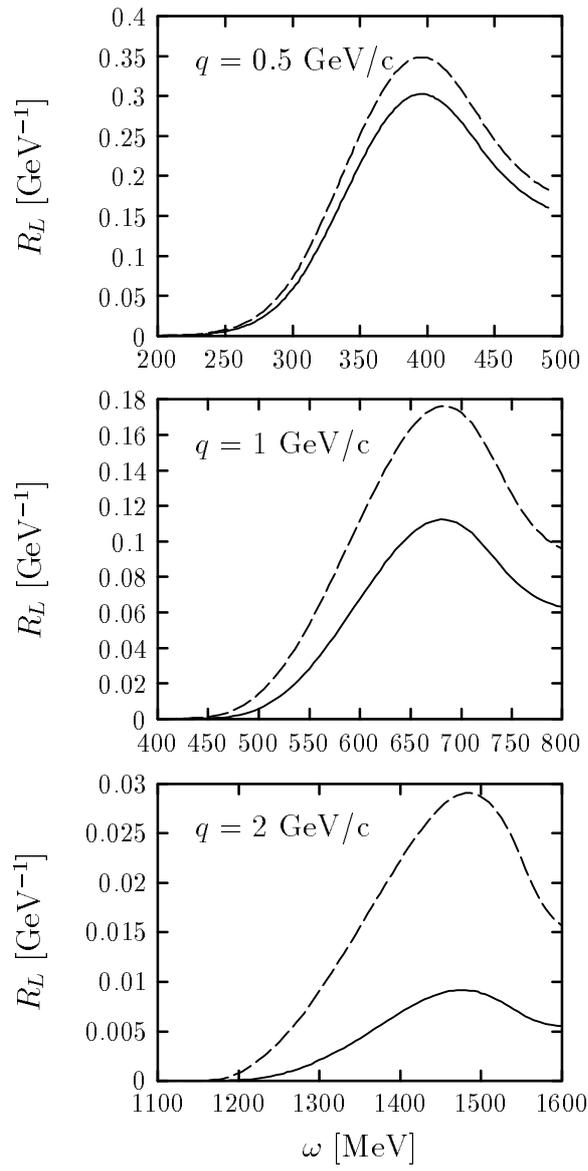
{\bf Figure 6:} Longitudinal response
for $^{12}$C. 
 Solid: $\Delta$-peak  without Coulomb form factor; 
dashed: $\Delta$-peak with Coulomb form factor 
$G_C(0)=-0.15G_M(0)$.

\end{document}